\documentclass[final,3p,times]{elsarticle}

\usepackage{lineno,hyperref}
\usepackage{amsmath}
\usepackage{amsfonts}
\usepackage{verbatim}

\usepackage{mdframed}
\definecolor{framecolor}{RGB}{255,230,210}

\mdfdefinestyle{MyFrame}{%
    linecolor=white,
    outerlinewidth=0pt,
    innertopmargin=4pt,
    innerbottommargin=4pt,
    innerrightmargin=4pt,
    innerleftmargin=4pt,
        leftmargin = 4pt,
        rightmargin = 4pt,
    skipabove=4pt,skipbelow=4pt,
    align=left,
    backgroundcolor=framecolor}
 
\newcommand{\eq}[1]{eq.~\eqref{eq:#1}}

\renewcommand{\sec}[1]{section~\ref{#1}}

\newcommand{\app}[1]{\ref{#1}}
\newcommand{\fig}[1]{figure~\ref{fig:#1}}

\newcommand{\Tab}[1]{table~\ref{tab:#1}}
\newcommand{\mycites}[1]{refs.~\cite{#1}}
\newcommand{\mycite}[1]{ref.~\cite{#1}}

\newcommand{\CP}{$\mathcal{CP}$}

\newcommand{\Zsym}{$\mathbb{Z}_2$ }
\newcommand{\real}[1]{\mathrm{Re}\left(#1\right)}
\newcommand{\imag}[1]{\mathrm{Im}\left(#1\right)}

\newcommand{\nn}{\nonumber\\}
\newcommand{\abs}[1]{\lvert#1\rvert}

\newcommand{\braket}[1]{\left\langle #1\right\rangle}

\newcommand{\code}[1]{\texttt{#1}}
\newcommand{\file}[1]{\textit{#1}}
\newcommand{\THDM}{\code{THDM}}
\newcommand{\SM}{\code{SM}}
\newcommand{\THDME}{\code{2HDME}}

\newcounter{bla}

\journal{Computer Physics Communications}

\begin{document}

\begin{frontmatter}

  \title{\code{2HDME}: Two-Higgs-Doublet Model Evolver}

  \author{Joel Oredsson}
  
  \ead{joel.oredsson@thep.lu.se}
  \address{Department of Astronomy and Theoretical Physics, Lund University,\\
   S\"olvegatan 14A, SE 223-62 Lund, Sweden}
  
  \begin{abstract}
  Two-Higgs-Doublet Model Evolver (\code{2HDME}) is a \code{C++} program that provides the functionality to perform fast renormalization group equation running of the general, potentially \CP-violating, 2 Higgs Doublet Model at 2-loop order.
  Simple tree-level calculations of masses; calculations of the oblique parameters $S$, $T$ and $U$; different parameterizations of the scalar potential; tests of perturbativity, unitarity and tree-level stability of the scalar potential are also implemented. 
  We briefly describe the \THDME's structure, provide a demonstration of how to use it and list some of the most useful functions.\\
  
  \end{abstract}
  
  
  \end{frontmatter}

{\bf PROGRAM SUMMARY}
%Delete as appropriate.

\begin{small}
\noindent
{\em Program Title: 2HDME}                                          \\
{\em Licensing provisions: GNU GPLv3 }                                   \\
{\em Programming language: C++}                                   \\
{\em Nature of problem: Renormalization group evolution of the general, potentially complex, 2 Higgs doublet model at 2-loop order. Also tests of perturbativity, unitarity and tree-level stability of the model.}\\
{\em Solution method: Numerical solutions of systems of ordinary differential equations.}

\end{small}

%%%%%%%%%%%%%%%%%%%%%%%%%%%%%%%%%%%%%%%%%%%%%%%%%%%%%%%%%%%%%%%%%%%%%%%%%%%%%%%% 
\section{Introduction}\label{intro}
%%%%%%%%%%%%%%%%%%%%%%%%%%%%%%%%%%%%%%%%%%%%%%%%%%%%%%%%%%%%%%%%%%%%%%%%%%%%%%%%

The discovery of a 125 GeV scalar particle at the LHC \cite{Aad:2012tfa, Chatrchyan:2012xdj} marks the beginning of an era of precision Higgs physics.
So far, it resembles the Higgs boson of the Standard Model (SM) \cite{Khachatryan:2016vau}; however, further experimental investigation is required to decipher its true nature.

The Two-Higgs-Doublet Model (2HDM) is a very popular extension of the standard model.
By adding a second Higgs doublet, it offers a rich phenomenology with three neutral and one charged pair of Higgs bosons.
Some of its features are the possibility of explicit and spontaneous \CP-violation in the scalar sector; a dark matter candidate and lepton number violation.
For a review of the 2HDM, we refer to \mycite{Branco:2011iw}.

A useful tool when investigating the 2HDM is to employ a renormalization group (RG) analysis.
One can with such a method look for instabilities, fine-tuning and valid energy ranges in the 2HDM's parameter space.
The RG equations (RGEs) for any renormalizable quantum field theory at 2-loop order are well known \cite{Machacek:1983tz, Machacek:1983fi, Machacek:1984zw,Luo:2002ti}.
However, implementing them for a specific model to perform numerical calculations is a tedious and error-prone task.

The purpose of 2 Higgs Doublet Model Evolver (\THDME) is to provide a fast \code{C++} application programming interface (API) that can be used to evolve the 2HDM in renormalization scale by numerically solving its RGEs.
\THDME~works with the general, potentially complex and \CP-violating, 2HDM and both 1- and 2-loop RGEs are implemented.
Furthermore, calculations of oblique parameters, tests of perturbativity, unitarity and tree-level stability of the scalar potential are available.

In this manual, we give instructions and showcase some of \THDME's functionalities.
The source code can be found at \mycite{2HDME} and we give some installation instructions in \app{installation}.
For a physics discussion, we refer to \mycite{Oredsson:2018yho}; which employs \THDME~to analyze \Zsym breaking effects in the evolution of the 2HDM.

We begin by briefly describing the structure of \THDME~in \sec{2HDMEstruct} and a demonstration of how to use the API is given in \sec{demo}; installation instructions are given in \app{installation}.
Further details on the base classes and the functionality they provide are then given in \sec{classes}.
The main class of \THDME~is \THDM~which is described in \sec{THDM}, where we also give a short review of the physics of the 2HDM.
We give a short description of the algorithm used when performing RG evolution in \sec{RGEalgorithm}.
Instructions of how to implement one's own model or extend the \THDM~class are given in \sec{extension}.
Finally, we discuss other software that are capable of performing RG evolution in \sec{comparison} and conclude in \sec{conclusion}.

%%%%%%%%%%%%%%%%%%%%%%%%%%%%%%%%%%%%%%%%%%%%%%%%%%%%%%%%%%%%%%%%%%%%%%%%%%%%%%%%
\section{Structure of 2HDME}\label{2HDMEstruct}
%%%%%%%%%%%%%%%%%%%%%%%%%%%%%%%%%%%%%%%%%%%%%%%%%%%%%%%%%%%%%%%%%%%%%%%%%%%%%%%%

\THDME~is written in \code{C++11} and depends on GSL \cite{GSL} for numerically solving the RGEs as well as on Eigen \cite{eigenweb} for linear algebra operations, see \app{installation} for installation instructions.
All source code is fairly well documented with comments in the header files, which show the functionality of all the classes.
\THDME~originated from an extension of \code{2HDMC} \cite{Eriksson:2009ws} and hence share a similar structure.

The purpose of \THDME~is to provide an API that consists of methods to manipulate a 2HDM model; thus
the idea is that the user should write their own executable code that uses the \code{THDM} class of \THDME.
A simple example that demonstrates how to use \THDME~is provided in \file{src/demos/DemoRGE.cpp}; which is explained in more detail in \sec{demo}.

The main class of the \THDME~is the \code{THDM} object that describes a general, potentially complex, 2HDM; see \sec{THDM} for a more detailed description of its functionality.
For basic usage, one only needs to interact with the public methods of \code{THDM} (and the \code{SM} class to set boundary conditions).
The framework to solve RGEs is contained in the \code{RgeModel} class, which \code{THDM} inherits from.
The \code{RgeModel}, furthermore, inherits some basic functionality related to data and console output from the abstract class \code{BaseModel}.
To initialize a THDM object, one needs to specify three sectors: the Standard Model (SM) parameters, the scalar potential and the Yukawa sector.

The SM parameters that need to be set are the Higgs vacuum expectation value, fermion masses, CKM matrix, gauge couplings and renormalization scale at which they are defined.
This can be done manually with a member function, where the user specifies all the input.
Another method is to initialize the \THDM~class using the \SM~class that represents the SM.
The \SM~is by default initialized at the top mass scale, $\approx$173 GeV, see \app{SMinput} for a detailed list of all the values; however, these can all be easily changed using its member functions.
Similar to the \THDM, the \SM~also inherits RGE functionality from the \code{RgeModel} class, with its own set of RGEs.
Thus, if one wants the SM parameters at another renormalization scale, the \SM~can be evolved to an arbitrary energy scale.
However, one should only run models above the top mass scale, since the RGEs specified for the SM include all particles\footnote{Running the SM downwards in energy below the top mass scale should incorporate some mechanism for integrating out particles at their corresponding mass threshold. This is currently not implemented in \THDME.}.

After setting up all the SM parameters, the \THDM~needs a scalar potential and a Yukawa sector.
The scalar sector can be specified with any of the 2HDM bases in \sec{bases};
these bases are separate \code{struct}s that are defined in \file{THDM\_bases.cpp}.   
Note though that \THDM~internally works in the generic basis.

The Yukawa sector can be specified in three ways.
One option is to impose a \Zsym symmetry; type-I,-II,-III(Y) or -IV(X); these are defined in \Tab{Z2symmetries}.
Another is to use a flavor alignment ansatz, where the Yukawa matrices for the different Higgs fields are proportional to each other.
Of course one can also set the Yukawa matrices manually as a third option.

To evolve the \THDM~or \SM, one simply uses their member function \code{evolve()}.
The options for the RG evolution can be set with the \code{RgeConfig} \code{struct}.

%%%%%%%%%%%%%%%%%%%%%%%%%%%%%%%%%%%%%%%%%%%%%%%%%%%%%%%%%%%%%%%%%%%%%%%%%%%%%%%%
\subsection{2-loop RGEs}\label{rges}
%%%%%%%%%%%%%%%%%%%%%%%%%%%%%%%%%%%%%%%%%%%%%%%%%%%%%%%%%%%%%%%%%%%%%%%%%%%%%%%%

The 2-loop RG equations (RGEs) for massless parameters of any renormalizable quantum field theory in 4 dimensions were derived in the seminal papers by Machacek and Vaughn \cite{Machacek:1983tz, Machacek:1983fi, Machacek:1984zw}.
That work has been supplemented with the 2-loop RGEs of massive parameters in \mycite{Luo:2002ti}, which is the source that we have used to derive the 1- and 2-loop RGEs for a general 2HDM.  
Note though, that when working with quantum field theories with multiple indistinguishable scalar fields one must be careful when interpreting their formulas, since the formulas in  \mycite{Machacek:1983tz,Machacek:1983fi,Machacek:1984zw,Luo:2002ti} are written for the case of an irreducible representation of the scalar fields
\footnote{This is a subtlety that is also discussed in \mycite{Bednyakov:2018cmx, Schienbein:2018fsw}.}.
In the case of a general 2HDM, one gets non-diagonal anomalous dimensions that mixes the scalar fields during RG evolution\footnote{For more details about the RGEs of different renormalization schemes in theories with multiple indistinguishable scalar fields, we refer to \mycite{Bijnens:2018rqw}.}.

The general 2-loop RGEs are very long and we will thus not write them down here, but are instead provided as supplementary material and are of course also available in \code{C++} format in the source code of \THDME\footnote{They are collected in separate header files in \file{src/RGEs}.}.

%%%%%%%%%%%%%%%%%%%%%%%%%%%%%%%%%%%%%%%%%%%%%%%%%%%%%%%%%%%%%%%%%%%%%%%%%%%%%%%%
\section{Demonstration of usage}\label{demo}
%%%%%%%%%%%%%%%%%%%%%%%%%%%%%%%%%%%%%%%%%%%%%%%%%%%%%%%%%%%%%%%%%%%%%%%%%%%%%%%%

As an example of how to use the \THDME~API, we here go through the \file{DemoRGE} in \file{src/demos}, where a 2HDM is initialized at the top mass scale and then evolved up to the Planck scale.
For instructions of how to install and run the demo, see \app{installation}.

The first thing to note is that the \THDME~is wrapped in a \code{namespace}, thus including \code{using namespace THDME} is convenient.

To initialize the \THDM, we first need to specify the SM parameters.
Rather then specifying them manually, we can use the \SM~class to provide all the necessary inputs.
It is constructed at the top mass scale and we print its parameters to the console with
\begin{mdframed}[style=MyFrame,userdefinedwidth=20em]
\code{SM sm; \\ sm.print\_all();}
\end{mdframed}
This \SM~is used to initialize the CKM matrix, fermion mass $\kappa^F$ matrices, gauge couplings, vacuum expectation value (VEV) $v$ and the renormalization scale of the \THDM.
For instructions of how to obtain a \SM~at another renormalization scale, see \sec{SM}.
Note, that the \SM~parameters can also be set manually with member functions that are described in \sec{manualSM}. 

To create a \THDM~and feed it a \SM, use
\begin{mdframed}[style=MyFrame,userdefinedwidth=20em]
\code{THDM thdm(sm);}
\end{mdframed}
Next up, we need to set the scalar potential. 
This can be done with any of the bases in \sec{bases}.
The bases are defined as \code{struct} objects which have member functions that can be used to convert one base to another.
Here, we use the generic basis which we specify by\footnote{There is also the VEV phase $\xi$ which is automatically initialized to zero; furthermore, it is fixed by the tadpole equations when actually setting the \THDM~potential.}
\begin{mdframed}[style=MyFrame,userdefinedwidth=30em]
\code{Base\_generic gen;                       \\
  gen.beta = 1.46713;\\
  gen.M12 = std::complex<double>(3132.85, 0.);\\
  gen.Lambda1 = 0.413702;\\
  gen.Lambda2 = 0.263926;\\
  gen.Lambda3 = 0.13313;\\
  gen.Lambda4 = -0.0444794;\\
  gen.Lambda5 = std::complex<double>(0.29586, 0.);\\
  gen.Lambda6 = std::complex<double>(0., 0.);\\
  gen.Lambda7 = std::complex<double>(0., 0.);
 }
\end{mdframed}
We set the potential with
\begin{mdframed}[style=MyFrame,userdefinedwidth=20em]
\code{thdm.set\_param\_gen(gen);}
\end{mdframed}
Finally we set the Yukawa sector to be type I \Zsym symmetric with
\begin{mdframed}[style=MyFrame,userdefinedwidth=20em]
\code{thdm.set\_yukawa\_type(TYPE\_I);}
\end{mdframed}

Now that we have fully initialized the \THDM~and the \SM, we can save them in SLHA-like text files with
\begin{mdframed}[style=MyFrame,userdefinedwidth=20em]
\code{sm.write\_slha\_file(); \\ thdm.write\_slha\_file();}
\end{mdframed}

\begin{figure}[h!]
  \begin{center}
\includegraphics[trim=0cm 0cm 0cm 0cm,clip,width=12cm]{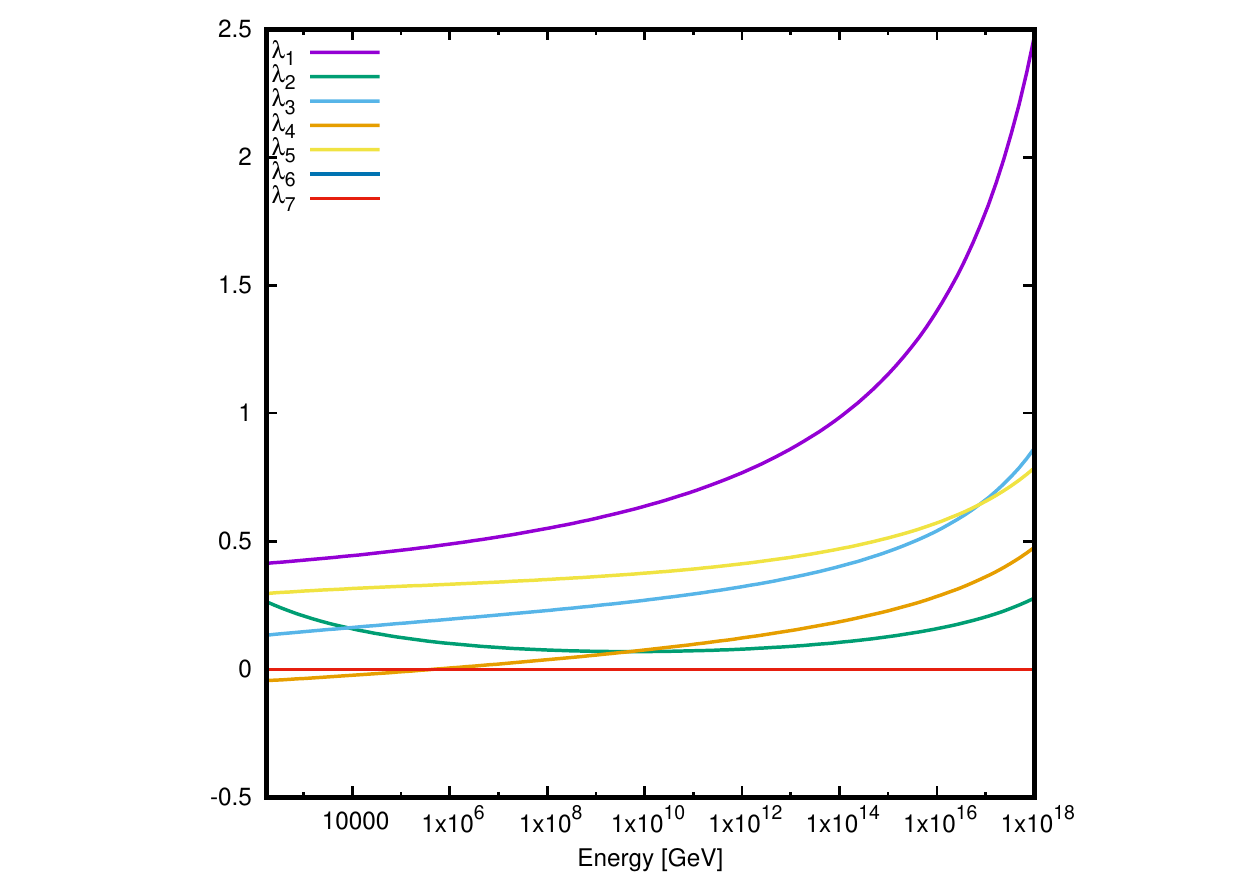}
\caption{The evolution of quartic couplings in the generic basis produced by \file{DemoRGE.cpp}.}
\label{fig:demoEx}
  \end{center}
\end{figure}

To evolve the \THDM, we need to configure the settings of the RG evolution.
This is done by creating a \code{RgeConfig} \code{struct} and feeding it to the \THDM~like
\begin{mdframed}[style=MyFrame,userdefinedwidth=25em]
\code{RgeConfig options; \\
  options.dataOutput = true;         \\
  options.consoleOutput = true;      \\
  options.evolutionName = "DemoRGE"; \\
  options.twoloop = true;            \\
  options.perturbativity = true;     \\
  options.stability = false;         \\
  options.unitarity = false;         \\
  options.finalEnergyScale = 1e18; \\
  options.steps = 100;\\ 
  thdm.set\_rgeConfig(options);
}
\end{mdframed}
The different options are explained in \sec{RgeModel}.
One can print the options to the console with \code{options.print()}.
Now, we are ready to evolve the \THDM~with
\begin{mdframed}[style=MyFrame,userdefinedwidth=20em]
\code{thdm.evolve();}
\end{mdframed}
which should only take a couple of seconds at 2-loop order.
The parameters as functions of the renormalization scale are listed in \file{output/DemoRGE/data} and basic plots are created in \file{output/DemoRGE/plots}, see \fig{demoEx} for an example.

Afterwards we can print the parameters at the energy scale where the RG running stopped with 
\begin{mdframed}[style=MyFrame,userdefinedwidth=20em]
\code{thdm.print\_all();}.
\end{mdframed}
To retrieve the scalar potential, one can for example use 
\begin{mdframed}[style=MyFrame,userdefinedwidth=20em]
  \code{thdm.get\_param\_gen();}
\end{mdframed}

Next up, we can evolve the \THDM~to another energy.
First we save the \THDM~at the high scale with
\begin{mdframed}[style=MyFrame,userdefinedwidth=35em]
  \code{thdm.write\_slha\_file(sphenoLoopOrder, "DemoRGE\_evolvedThdm");}.
\end{mdframed}
It is always possible to evolve the \code{thdm} further.
After the fist evolution, the \THDM~is specified at the high scale and to evolve the it downwards to $\mu=1$ TeV, all one has to do is to change the \code{finalEnergyScale} of its \code{RgeConfig} before evolving, which is achieved with
\begin{mdframed}[style=MyFrame,userdefinedwidth=20em]
  \code{options.finalEnergyScale = 1e3;}\\
  \code{options.dataOutput = false;}\\
  \code{thdm.set\_rgeConfig( options);}\\
  \code{thdm.evolve();}
\end{mdframed}
where we also changed \code{dataOutput} to \code{false}; thus preventing that the first plots are overwritten.
A more compact method would be to use \code{thdm.evolve\_to(1e3)}, which sets the final scale to the given argument.

%%%%%%%%%%%%%%%%%%%%%%%%%%%%%%%%%%%%%%%%%%%%%%%%%%%%%%%%%%%%%%%%%%%%%%%%%%%%%%%%
\section{Classes}\label{classes}
%%%%%%%%%%%%%%%%%%%%%%%%%%%%%%%%%%%%%%%%%%%%%%%%%%%%%%%%%%%%%%%%%%%%%%%%%%%%%%%%

There are two main classes of \THDME: \THDM~and \SM.
These inherit the features from the base classes \code{BaseModel} and \code{RgeModel} like
\begin{align*}
  \text{\code{BaseModel}}\leftarrow~& \text{\code{RgeModel}}\leftarrow \text{\THDM},\nn
  \text{\code{BaseModel}}\leftarrow~& \text{\code{RgeModel}}\leftarrow \text{\SM}.
\end{align*}
We briefly describe the features of the \code{BaseModel}, \code{RgeModel} and \SM~in the following subsections, while the \THDM~is described in more detail in \sec{THDM}.

%%%%%%%%%%%%%%%%%%%%%%%%%%%%%%%%%%%%%%%%%%%%%%%%%%%%%%%%%%%%%%%%%%%%%%%%%%%%%%%%
\subsection{\code{BaseModel} class}\label{BaseModel}
%%%%%%%%%%%%%%%%%%%%%%%%%%%%%%%%%%%%%%%%%%%%%%%%%%%%%%%%%%%%%%%%%%%%%%%%%%%%%%%%

The \code{BaseModel} is an abstract class that offers some basic functionality such as input and output to data files as well as to the console.

The level of information printed to the console of the \THDM~ and \SM~during computations can be set by 
\begin{center}
  \begin{tabular}{l|l|c}
    Function & Description & Returns \\ 
    \hline
    \code{set\_logLevel(LogLevel)} & Sets level of console output information & \code{void} \\
  \end{tabular}
\end{center}
where \code{LogLevel} can be either one of the following:
\begin{itemize}
  \item \code{LOG\_INFO}: Prints information of calculations performed and status updates.
  \item \code{LOG\_ERRORS}: Only prints error messages.
  \item \code{LOG\_WARNINGS}: Prints error messages and warning messages.
  \item \code{LOG\_DEBUG}: Prints all the information as well as additional debugging information.
\end{itemize}

%%%%%%%%%%%%%%%%%%%%%%%%%%%%%%%%%%%%%%%%%%%%%%%%%%%%%%%%%%%%%%%%%%%%%%%%%%%%%%%%
\subsection{\code{RgeModel} class}\label{RgeModel}
%%%%%%%%%%%%%%%%%%%%%%%%%%%%%%%%%%%%%%%%%%%%%%%%%%%%%%%%%%%%%%%%%%%%%%%%%%%%%%%%

The \code{RgeModel} inherits the input/output functionality of the \code{BaseModel} and acts as a base class which offers a framework to incorporate RG evolution in derived classes;
both \THDM~and \SM~are derived classes of \code{RgeModel}.

If one wants to create a new type of class that implements RGE functionality similar to \THDM~and \SM, it is easy to use \code{RgeModel}. 
For example one might want to extend the 2HDM with additional operators and consequently new parameters and RGEs.
Instructions of how to do this is given in \sec{extension}.
There is also the class \code{NewModel}, that serves as a minimalistic example of a new class that inherits from \code{RgeModel}.

See the header file \code{RgeModel.h} for a list of all member functions.
Some of the most important ones are:
\begin{center}
  \begin{tabular}{l|l|c}
    Function & Description & Returns \\ 
    \hline
    \code{set\_rgeConfig(RgeConfig)} & Sets up the options for the RG evolution. & \code{void} \\
    \code{get\_rgeConfig()} & Retrieves the RG evolution options. & \code{RgeConfig} 
  \end{tabular}
\end{center}

The options for performing RG evolution are saved in the \code{RgeConfig} member variable \code{\_rgeConfig}\footnote{All member variables of classes are denoted with an underline as a first character.}. 
This \code{RgeConfig} has the following member variables:
\begin{itemize}
	\item \code{bool twoloop}: If true, uses 2-loop RGEs; otherwise uses 1-loop.
    \item \code{bool perturbativity}: If true, the RG evolution stops when perturbativity is violated.
    \item \code{bool unitarity}: If true, the RG evolution stops when unitarity is violated.
    \item \code{bool stability}: If true, the RG evolution stops when tree-level stability is violated.
    \item \code{bool consoleOutput}: If true, prints information to the console during and after RG evolution.
    \item \code{string evolutionName}: Name of folder in \file{output}, where data and plots are stored.
    \item \code{bool dataOutput}: If true, creates data files in \file{output/"evolutionName"/data} folder. These files contain the parameters of the model at each step in the RG evolution. If \code{GNUPLOT} is enabled in the \file{Makefile}, simple plots are created in \file{output/"evolutionName"/plots}.
    \item \code{double finalEnergyScale}: The final energy scale, in GeV, for the RG running (from current renormalization scale). This can be both higher as well as lower than the current renormalization scale.
    \item \code{int steps}: Number of steps for the RG evolution; which are logarithmically distributed. Perturbativity, unitarity and stability are being checked at each step.
\end{itemize}

To evolve a class that inherits from \code{RgeModel}, one can use either of the following two functions:
\begin{center}
  \begin{tabular}{l|c|c}
    Function & Description & Returns \\ 
    \hline
    \code{evolve()} & Evolves the model in renormalization scale. & \code{true/false}\\
    \code{evolve\_to(double)} & Evolves the model to given scale & \code{true/false}
  \end{tabular}
\end{center}
These evolve the model according to the configuration set by \code{set\_rgeConfig(RgeConfig)}.
It returns false if the ODE solver runs into numerical problems, \textit{e.g.}\ encounters a Landau pole.
This does not usually happen if \code{perturbativity} is set to \code{true} in the \code{RgeConfig}, since the RG running is stopped before the parameters become too large.

The function \code{evolve\_to} first sets \code{\_rgeConfig.finalEnergyScale} to the given argument and then evolves the model.

The result of the RG evolution is collected in a \code{RgeResults} \code{struct}.
It can be retrieved with \code{get\_rgeResults()} or simply printed to the console with \code{print\_rgeResults()}.  
It stores any violation of perturbativity, unitarity or stability and at what energy scale it occurs.

Since the \code{evolve()} function updates all the parameters, it can be useful to save the state of a model at a specific renormalization scale using
\begin{center}
  \begin{tabular}{l|c|c}
    Function & Description & Returns \\ 
    \hline
    \code{save\_current\_state()} & Saves the current state internally & \code{void}\\
    \code{reset\_to\_saved\_state()} & Resets to a previously saved state & \code{true/false}
  \end{tabular}
\end{center}
The model can afterwards be restored to this state with \code{reset\_to\_saved\_state()}.
There is however currently no way of saving multiple states. 
If one wishes to do so, it might be easiest to simply copy \code{THDM}~objects instead.

Some other useful functions are:
\begin{center}
  \begin{tabular}{l|l|c}
    Function & Description & Returns \\ 
    \hline
    \code{set\_final\_energy\_scale(double)} & Sets \code{finalEnergyScale} for RG evolution & \code{void}\\
    \code{get\_rgeResults(RgeResults)} & Retrieves results of RG evolution \ & \code{RgeResults}\\
    \code{print\_rgeResults(RgeResults)} & Prints results of RG evolution to console \ & \code{void}\\
    \code{set\_renormalization\_scale()} & Sets $\mu$ & \code{void} \\
    \code{get\_renormalization\_scale()} & Retrieves $\mu$ & \code{double} 
  \end{tabular}
\end{center}

%%%%%%%%%%%%%%%%%%%%%%%%%%%%%%%%%%%%%%%%%%%%%%%%%%%%%%%%%%%%%%%%%%%%%%%%%%%%%%%%
\subsection{\code{SM} class}\label{SM}
%%%%%%%%%%%%%%%%%%%%%%%%%%%%%%%%%%%%%%%%%%%%%%%%%%%%%%%%%%%%%%%%%%%%%%%%%%%%%%%%

A class that describes the SM.
It inherits the RGE functionality from \code{RgeModel}.
The physics member variables that are evolved during RG evolution are
\begin{itemize}
	\item The three SU(3)$_c$, SU(2)$_L$ and U(1)$_Y$ gauge couplings: \code{\_g3}, \code{\_g2}, \code{\_g1}.
	\item Quartic Higgs couplings $\lambda$: \code{\_lambda}; normalized according to the scalar potential
  \begin{align}
    V = -\frac{m^2}{2}\Phi^\dagger \Phi + \lambda (\Phi^\dagger \Phi)^2.
  \end{align}
	\item Higgs VEV $v=(\sqrt{2}G_F)^{-1/2}\approx 246$ GeV: \code{\_v}.
	\item Complex 3 by 3 Yukawa matrices, \code{\_yU}, \code{\_yD} and \code{\_yL}.
\end{itemize}
In total these sum up to 59 real parameters.
The Yukawa matrices are in the fermion weak eigenbasis and initialized with the CKM matrix and fermion masses:
\begin{align}
	Y^U =~& \frac{\sqrt{2}}{v}M_u,\nn
	Y^D =~& \frac{\sqrt{2}}{v}V_{CKM}M_d,\nn
	Y^L =~& \frac{\sqrt{2}}{v}M_\ell,
\end{align}
where $M_f$ are the diagonal fermion mass matrices.
These parameters are at construction defined at the renormalization scale $\mu=173.34$ GeV.
For the default numerical values and conventions for the CKM matrix, see \app{SMinput}.
One can modify all the default values with the member functions in \sec{manualSM}.
There is no mechanism to generate neutrino masses implemented; hence the neutrinos are treated as being massless.

\subsubsection{Functionality}

The \SM~can be saved to a SLHA-like text file with
\begin{center}
  \begin{tabular}{l|l|c}
    Function & Description & Returns \\ 
    \hline
    \code{write\_slha\_file(string)} & Creates SLHA file. & \code{void} \\
    \code{set\_from\_slha\_file(string)} & Sets the \SM~from SLHA file. & \code{true/false} 
  \end{tabular}
\end{center}

Other useful functions are
\begin{center}
  \begin{tabular}{l|l|c}
    Function & Description & Returns \\ 
    \hline
    \code{get\_v2()} & Returns $v^2$ & \code{double} \\
    \code{get\_gauge\_couplings()} & Returns $\{g_1,g_2,g_3\}$ & \code{vector<double>} \\
    \code{get\_mup()} & Returns $\{m_u,m_c,m_t\}$ & \code{vector<double>} \\
    \code{get\_mdn()} & Returns $\{m_d,m_s,m_b\}$ & \code{vector<double>} \\
    \code{get\_ml()} & Returns $\{m_e,m_\mu,m_\tau\}$ & \code{vector<double>} \\
    \code{get\_vCkm()} & Returns CKM matrix & \code{Eigen::Matrix3cd} \\
    \code{get\_lambda()} & Returns Higgs quartic coupling & \code{double} \\
    \code{print\_all()} & Prints parameters to console & \code{void}
  \end{tabular}
\end{center}

\subsubsection{Changing renormalization scale}

As previously mentioned, the \SM~is constructed at the top mass scale.
It is however possible to obtain a \SM~defined at any other energy scale.
One should use the functions \code{evolve()} or \code{evolve\_to(double)} of \code{RgeModel} to evolve the \SM.
In the RG evolution, the mass matrices $M_f$ and CKM matrix are calculated at each step by diagonalizing the $Y^F$ matrices.
Note though that the RGEs for the SM are the full ones, with 6 quarks for example, and no decoupling is performed which should be done when running at energy scales that are below the top mass scale.
For example, to get the \SM~at 1 TeV, one can evolve a constructed \SM~with \code{evolve\_to(1e3)}.

\subsubsection{Setting the parameters}\label{manualSM}

As previously mentioned the \SM~is created with some default numerical values for its parameters.
Of course, these can also be set manually with the following functions:
\begin{center}
\begin{tabular}{l|cc}
  Function & Description \\ 
  \hline
  \code{set\_params(mu,lambda,v2,g\_i,mup,mdn,ml,VCKM)} &  Sets all parameters (\code{mu} = renormalization scale).  \\
  \code{set\_v2(v2)} &  Sets the squared VEV. \\
  \code{set\_higgs(v2, lambda)} &  Sets the potential parameters. \\
  \code{set\_gauge(g\_i)} &  Sets the gauge couplings from a \code{vector<double>}. \\
  \code{set\_mup(mup)} &  Sets the up quark masses from a \code{vector<double>}. \\
  \code{set\_mdn(mdn)} &  Sets the down quark masses from a \code{vector<double>}. \\
  \code{set\_ml(ml)} &  Sets the lepton masses from a \code{vector<double>}. \\
  \code{set\_ckm(VCKM)} &  Sets the CKM matrix from an \code{Eigen::Matrix3cd}.
\end{tabular}
\end{center}

Another method of setting the parameters manually is by loading a SLHA file.
This can be done as follows:
\begin{itemize}
  \item Construct a \SM~and save the parameters to a SLHA file with \code{write\_slha\_file("smSlhaFile")}.
  \item The SLHA file contain all the parameters of the \SM, including the renormalization scale. It is readable and thus provides an easy way to manually edit all the numerical values.
  \item Then use the edited SLHA file to set a \SM~object with \code{set\_from\_slha\_file("smSlhaFile")}.
\end{itemize}

%%%%%%%%%%%%%%%%%%%%%%%%%%%%%%%%%%%%%%%%%%%%%%%%%%%%%%%%%%%%%%%%%%%%%%%%%%%%%%%%
\section{\code{THDM} class}\label{THDM}
%%%%%%%%%%%%%%%%%%%%%%%%%%%%%%%%%%%%%%%%%%%%%%%%%%%%%%%%%%%%%%%%%%%%%%%%%%%%%%%%

\THDM~ is the main class of \THDME~and describes a general, potentially complex, 2HDM.
It inherits RGE functionality from \code{RgeModel}. 

Here, we give a short summary of the general 2HDM and the parameterization of it inside \THDM.
For a thorough review of the 2HDM see \mycite{Branco:2011iw}.
We use the notation of \mycites{Davidson:2005cw, Haber:2006ue, Haber:2010bw} to describe the generic and Higgs bases of the 2HDM.

\subsection{Parameters of 2HDM}

The 2HDM contains two hypercharge $+1$ complex scalar SU(2) doublets, $\Phi_1$ and $\Phi_2$.
First of all, since the scalar fields have identical quantum numbers, one can always perform a field redefinition of the scalar fields, i.e.\ a non-singular complex transformation $\Phi_a\to B_{a\bar{b}}\Phi_b$.
The Lagrangian of the 2HDM exhibits a U(2) Higgs flavor symmetry, $\Phi_a \rightarrow U_{a\bar{b}} \Phi_b$; since the Lagrangian keeps the same form after such a transformation.
We will denote 2HDMs related by such Higgs flavor transformations as different bases of the 2HDM.
All the different bases that are implemented in \THDME~are listed and described in \app{bases}.

\subsection{Generic basis}\label{genericBasis}

The most general 2HDM gauge invariant renormalizable scalar potential can be written
\begin{align}\label{eq:GenericPotential}
		-\mathcal{L}_V=&m_{11}^2\Phi_1^\dagger\Phi_{1}+m_{22}^2\Phi_2^\dagger\Phi_{2}-(m_{12}^2\Phi_1^\dagger\Phi_{2}+\text{h.c.})+\frac{1}{2}\lambda_1\left(\Phi_1^\dagger\Phi_{1}\right)^2+\frac{1}{2}\lambda_2\left(\Phi_2^\dagger\Phi_{2}\right)^2\nonumber\\
		&+\lambda_3\left(\Phi_1^\dagger\Phi_{1}\right)\left(\Phi_2^\dagger\Phi_2\right)+\lambda_4\left(\Phi_1^\dagger\Phi_{2}\right)\left(\Phi_2^\dagger\Phi_1\right)
		\nonumber\\
		&+\left[\frac{1}{2}\lambda_5\left(\Phi_1^\dagger\Phi_2\right)^2+\lambda_6\left(\Phi_1^\dagger\Phi_1\right)\left(\Phi_1^\dagger\Phi_2\right)+\lambda_7\left(\Phi_2^\dagger\Phi_2\right)\left(\Phi_1^\dagger\Phi_2\right)+\text{h.c.}\right],
\end{align}
where $m_{12}^2,~\lambda_5,~\lambda_6$ and $\lambda_7$ are potentially complex while all the other parameters are real; resulting in a total of 14 degrees of freedom.
However, three of these are fixed by the tadpole equations and one can be removed by a re-phasing of the second Higgs doublet.
The bases in \eq{GenericPotential} will be referred to as the generic basis; which is the internal basis used in the \THDM~class.

After electroweak symmetry breaking, SU(2)$\times $U(1)$_Y\rightarrow $U(1)$_{\text{em}}$, both of the scalar fields acquire VEVs, which can be expressed in terms of a unit vector in the Higgs flavor space
\begin{align}\label{eq:vev}
	\braket{\Phi_a} = \frac{v}{\sqrt{2}} \left( \begin{array}{c} 0 \\ \hat{v}_a \end{array} \right), && \hat{v}_a \equiv \left( \begin{array}{c} c_\beta \\ s_\beta e^{i\xi} \end{array}\right),
\end{align}
where the unit vector is normalized to $\hat{v}_{\bar{a}}^* \hat{v}_a=1$.
By convention, we take $0\leq \beta \leq \pi/2$ and $0\leq \xi\leq 2\pi$.
Here, we have used up all our gauge freedom, when setting the VEV in the lower component of the doublets with a SU(2) transformation and removing any phase in the $\Phi_1$ VEV with a U(1)$_Y$ transformation. 
We also define $\hat{w}_b \equiv \hat{v}_{\bar{a}}^*\epsilon_{ab}$, where $\epsilon_{12}=-\epsilon_{21}=1$ and $\epsilon_{11}=\epsilon_{22}=0$.
The angle $\beta$ can also be expressed as the ratio of the Higgs fields' vacuum expectation values,
\begin{align}
  \tan\beta = \abs{\braket{\Phi_2}}/\abs{\braket{\Phi_1}}.
\end{align}

The Yukawa interactions in the generic basis are
\begin{align}
	-\mathcal{L}_Y=&\bar{Q}_L^0\tilde{\Phi}_{\bar{a}}\eta_a^{U,0}U_R^0+\bar{Q}_L^0\Phi_a\eta_{\bar{a}}^{D,0}D_R^0 + \bar{L}_L^0\Phi_a\eta_{\bar{a}}^{L,0}E_R^0 + \text{h.c.}~,
\end{align}
where the left-handed fermion fields in the weak eigenbasis are
\begin{align}
	Q_L^0 \equiv \left( \begin{array}{c} U_L^0 \\ D_L^0 \end{array} \right), && L_L^0 \equiv \left( \begin{array}{c} \nu_L^0 \\ E_L^0 \end{array} \right)
\end{align}
and $\tilde{\Phi}\equiv i\sigma_2 \Phi^*$.

The 129 parameters of the 2HDM are stored as member variables in \THDM:
\begin{itemize}
  \item The SU(3)$_c\times $SU(2)$_W\times $U(1)$_Y$ gauge couplings: \code{\_g3}, \code{\_g2} and \code{\_g1} respectively.
  \item The Higgs VEV $v^2 = v_1^2 + v_2^2$: \code{\_v2}. This is initialized when feeding a \SM~to the \THDM.
  \item The potential parameters in the generic basis: \code{\_base\_generic};
  which also includes the angles $\beta$ and $\xi$. See \app{bases} for a detailed description.
  \item The 6 Yukawa matrices in the fermion weak eigenbasis: \code{\_eta1U}, \code{\_eta2U}, \code{\_eta1D}, \code{\_eta2D}, \code{\_eta1L}, \code{\_eta2L}.  
\end{itemize}
These are the variables that have their RGEs defined in \file{RGE.cpp}.
Note though that the angles $\beta$ and $\xi$ are calculated from the VEVs of the Higgs fields, $v_a = v \hat{v}_a$; which run according to the anomalous dimensions of the fields.

In addition to the member variables above, the \THDM~also stores the Yukawa sector in the fermion mass eigenbasis.
To go to the fermion mass eigenbasis, the \THDM~first calculates the Yukawa matrices in the Higgs basis; which has the Lagrangian
\begin{align}
	-\mathcal{L}_Y =~& \bar{Q}_L^0 \tilde{H}_1 \kappa^{U,0} U_R^0 + \bar{Q}_L^0 H_1 \kappa^{D,0} D_R + \bar{L}_L^0 H_1 \kappa^{L,0} E_R^0\nn
	&+ \bar{Q}_L^0 \tilde{H}_2 \rho^{U,0} U_R^0 + \bar{Q}_L^0 H_2 \rho^{D,0} D_R + \bar{L}_L^0 H_2 \rho^{L,0} E_R^0 + \text{h.c.},
\end{align}
where only $H_1$ acquires a VEV.
The $\kappa^{F,0}$ and $\rho^{F,0}$ matrices are given by
\begin{align}
	\kappa^{U,0} =~& \hat{v}_a^*\eta_a^{U,0}, & \rho^{U,0} =& \hat{w}_a^*\eta_a^{U,0},\nn
	\kappa^{D,0} =~& \hat{v}_a\eta_a^{D,0}, &   \rho^{D,0} =& \hat{w}_a\eta_a^{D,0},\nn 
	\kappa^{L,0} =~& \hat{v}_a\eta_a^{L,0}, &   \rho^{L,0} =& \hat{w}_a\eta_a^{L,0}.
\end{align}
Note that $\hat{v}_a$ and $\hat{w}$ are defined in terms of the VEVs in the generic basis,
which run during RG evolution; thus the transformation to the Higgs basis is $\mu$-dependent.

After going to the Higgs basis, \THDM~performs biunitary transformations to diagonalize the $\kappa^{F,0}$ matrices.
The fermions in the mass eigenbasis are defined as
\begin{align}
	F_L \equiv V_L^F F_L^0, && F_R \equiv V_R^F F_R^0,
\end{align}
where $F\in \{U,D,E\}$ is each fermion species.
The diagonal Yukawa matrices are
\begin{align}
	\kappa^U =~& V_L^U\kappa^{U,0}V_R^{U\dagger}=\frac{\sqrt{2}}{v}\text{diag}(m_u,m_c,m_t),\nn
	\kappa^D =~& V_L^D\kappa^{D,0}V_R^{D\dagger}=\frac{\sqrt{2}}{v}\text{diag}(m_d,m_s,m_b),\nn
	\kappa^L =~& V_L^L\kappa^{L,0}V_R^{L\dagger}=\frac{\sqrt{2}}{v}\text{diag}(m_e,m_\mu,m_\tau),
\end{align}
while $\rho^{F} = V_L^F\rho^{F,0}V_R^{F\dagger}$ are potentially non-diagonal; which would mean that tree-level FCNCs are present.
The CKM matrix is composed out of the left-handed transformation matrices, $V_{CKM} \equiv V_L^UV_L^{D\dagger}$.

To summarize, in addition to the parameters in the generic basis, the \THDM~stores
\begin{itemize}
  \item The diagonal $\kappa^F$ matrices: \code{\_kU}, \code{\_kD} and \code{\_kL}. The fermion masses are also stored as \code{\_mup[3]}, \code{\_mdn[3]} and \code{\_ml[3]}.
  \item The non-diagonal $\rho^F$ matrices:  \code{\_rU}, \code{\_rD} and \code{\_rL}.
  \item The CKM matrix: \code{\_VCKM}
\end{itemize}

\subsection{How to use \THDM}

To fully initialize the \THDM, one needs to do three things in order: 
\begin{itemize}
  \item Construct a \THDM~object and either feed it a \SM~object or set the \SM~parameters manually.
  \item Set the scalar potential with any of the available bases.
  \item Fix the Yukawa sector.
        This can be done with a \Zsym symmetry or a flavor ansatz, which produces a Yukawa sector without FCNCs.
        However, the Yukawa matrices can also be set manually.
\end{itemize}.

\subsection{Setting the SM parameters}

The \THDM~needs the gauge couplings, CKM matrix, fermion masses and the square of the VEV before setting up the scalar potential and Yukawa sector.
This can be done manually by specifying all of these parameters, or through tree-level matching to the SM.

The \SM~can be given at construction or with \code{set\_sm(SM)}.
This will set the VEV, gauge couplings, renormalization scale, $\kappa^F$ matrices and CKM matrix.
A more sophisticated matching procedure than tree-level is beyond the scope of this program.
There are however physics scenarios where higher order matching, of at least the most important parameters, can have a very large effect \cite{Braathen:2017jvs}.
In such a scenario the user should derive all the \THDM's parameters in whatever matching scheme of their choosing.
Afterwards, these can be fed into the \THDM~with the following commands (all returning \code{void}):
\begin{center}
\begin{tabular}{l|cc}
  Function & Description \\ 
  \hline
  \code{set\_sm(mu,v2,g\_i,mup,mdn,ml,VCKM)} &  Sets all the necessary SM parameters (\code{mu} = renormalization scale). \\
  \code{set\_v2(v2)} &  Sets the squared VEV. \\
  \code{set\_gauge\_couplings(g\_i)} &  Sets the gauge couplings (\code{g\_i=vector<double>}). \\
  \code{set\_fermion\_masses(mup,mdn,ml)} &  Sets the fermion masses from \code{vector<double>} types. \\
  \code{set\_vCkm(VCKM)} &  Sets the CKM matrix from an \code{Eigen::Matrix3cd}.
\end{tabular}
\end{center}

\subsection{Setting the scalar potential}

The scalar potential can be set with any of the bases described in \app{bases}.
Internally though, the \THDM~uses the generic basis.
The functions are
\begin{center}
\begin{tabular}{l|cc}
  & Returns \\ 
  \hline
  \code{set\_param\_gen(Base\_generic,bool=true)} & \code{true/false} \\
  \code{set\_param\_higgs(Base\_higgs,bool=true)} & \code{true/false} \\
  \code{set\_param\_invariant(Base\_invariant,bool=true)} & \code{true/false} \\
  \code{set\_param\_hybrid(Base\_hybrid,bool=true)} & \code{true/false} \\
\end{tabular}
\end{center}
All of the functions set the parameters in the generic basis after transforming the basis that is given.
The optional \code{bool} argument refers to if the tree-level tadpole equations should be enforced; which they are by default.
If $\tan\beta\neq 0$, the eqs.(A4,A5,A7) of \mycite{Davidson:2005cw} are used, which fix $m_{11}^2$, $m_{22}^2$ and $\xi$. 
Otherwise the Higgs basis tadpole equations are used, which fix $m_{11}^2 = -v^2 \lambda_1/2$ and $m_{12}^2=-v^2 \lambda_6 /2$.
These functions will return \code{false} if the tree-level Higgs masses are imaginary or if the tadpole equations could not be set.

\subsection{Setting the Yukawa sector}\label{setYuk}

Note that $\tan\beta$ must be set before fixing the Yukawa sector.
After that has been done, it can initialized with 
\begin{center}
  \begin{tabular}{l|cc}
    & Description \\ 
    \hline
    \code{set\_yukawa\_type(Z2symmetry)} & Fixes all Yukawa matrices from a \Zsym symmetry.\\
    \code{set\_yukawa\_aligned(double,double,double)} & Sets the Yukawa matrices from a flavor ansatz.\\
    \code{set\_yukawa\_manual(Matrix3cd, \ldots)} &  Sets the $\rho^F$ matrices manually.\\
    \code{set\_yukawa\_eta(Matrix3cd, \ldots)} & Sets all $\eta^{F,0}$ manually. 
  \end{tabular}
  \end{center}
  The type of \Zsym symmetry is specified by a \code{enum}, \code{Z2symmetry}, which can be set to either \code{NO\_SYMMETRY}\footnote{Using \code{set\_yukawa\_type(NO\_SYMMETRY)} does nothing in terms of fixing the $\rho^F$ matrices.}, \code{TYPE\_I}, \code{TYPE\_II}, \code{TYPE\_III} or \code{TYPE\_IV}.
  Imposing a \Zsym symmetry makes the Yukawa matrices proportional to each other,
  \begin{align}
    \rho^F = a^F \kappa^F,
  \end{align}
  where the coefficients $a^F$ are fixed by $\beta$ as in \Tab{Z2symmetries}.
  These $a^F$ coefficients can also be set manually with \\\code{set\_yukawa\_aligned(aU,aD,aL)}.

  \begin{table}[h!]
		\centering
    		\begin{tabular}{|c|cccccc|}\hline
		Type	 & $U_R$	& $D_R$	& $L_R$ & $a^U$ & $a^D$ & $a^L$\\
		\hline 
		I & + & + & + &$\cot\beta$ & $\cot\beta$ & $\cot\beta$\\
		II & + & $-$ & $-$ &$\cot\beta$ & $-\tan\beta$ & $-\tan\beta$\\
		Y/III & + & $-$ & + &$\cot\beta$ & $-\tan\beta$ & $\cot\beta$\\
		X/IV & + & + & $-$ &$\cot\beta$ & $\cot\beta$ & $-\tan\beta$\\ \hline
		\end{tabular}
		\caption{Different \Zsym symmetries that can be imposed on the 2HDM. $\Phi_1$ is odd($-1$) and $\Phi_2$ is even($+1$). For every type of \Zsym symmetry, the $\rho^F$ matrices become proportional to the diagonal mass matrices, $\rho^F = a^F \kappa^F$. }
		\label{tab:Z2symmetries}
\end{table}

\subsection{Checks}

Some of the checks that are implemented are:
\begin{center}
  \begin{tabular}{l|cc}
    & Returns \\ 
    \hline
    \code{is\_perturbative()} & true/false\\
    \code{is\_unitary()} & true/false\\
    \code{is\_stable()} & true/false\\ 
    \code{is\_cp\_conserved()} & true/false
  \end{tabular}
\end{center}
These are simple tree-level constraints. 
It should be straightforward, however, to manually edit these functions and extend them with one's own algorithms if that is needed.

\subsubsection*{Perturbativity}

The perturbativity limit is reached when any of the $\lambda_i$ parameters is larger than a specific limit.
By default, this limit is set to $4\pi$; however, it can be changed with \code{set\_perturbativity\_limit(double)}.
The default value is somewhat arbitrary and should not be interpreted as a strict limit for when perturbation theory breaks down.
In fact, $4\pi$ is a very large when dealing with scalar quartic couplings.
When some $\lambda_i \sim \mathcal{O}(1)$ there are usually large quantum corrections to scalar masses for example and many tree-level quantities cannot be trusted.
Since the quartic couplings run very fast for $\lambda_i > 4\pi$, the perturbativity-violation scale can be interpreted as an approximation for the scale where a Landau pole is encountered.
Turning off the perturbativity limit is not recommended; since running a \THDM~above this limit produces numerical problems for the ODE solver and consequently slows down any evolution.

\subsubsection*{Perturbative unitarity}

Tree-level perturbative unitarity of the scattering matrix of scalars at large $\sqrt{s}$ produces constraints on the quartic couplings. 
These constraints are laid out in \app{unitarity}.
Note, that they are only valid for large scattering energies and there are scenarios where one needs to apply more careful checks \cite{Goodsell:2018fex}.

\subsubsection*{Tree-level stability of the scalar potential}

The stability conditions at tree-level are given in \app{stability}.

Similar to the above checks there are more sophisticated methods that can be used that includes quantum corrections. 
In \mycite{Staub:2017ktc} it is shown that by analyzing the 1-loop effective potential using Vevacious \cite{Camargo-Molina:2013qva}, one can in fact save many parameter points that are deemed unstable at tree-level.

\subsection{Miscellaneous functions}

There are a number of functions that return useful quantities from the \THDM:
\begin{center}
  \begin{tabular}{l|c}
    Function & Returns \\ 
    \hline
    \code{get\_param\_gen()} & \code{Base\_generic}\\
    \code{get\_param\_higgs()} & \code{Base\_higgs}\\
    \code{get\_param\_invariant()} & \code{Base\_invariant}\\
    \code{get\_param\_hybrid()} & \code{Base\_hybrid}\\
    \code{get\_yukawa\_type()} & \code{Z2symmetry}\\
    \code{get\_aF()} & $\{|a^U|, |a^D|, |a^L|\}$\\
    \code{get\_v2()} & $v^2$\\ 
    \code{get\_gauge\_couplings()} &  $\{g_1,g_2,g_3\}$\\
    \code{get\_mup()} & $\{m_{u}, m_c, m_t\}$\\
    \code{get\_ml()} & $\{m_e, m_\mu, m_\tau\}$\\
    \code{get\_vCkm()} & $V_{CKM}$\\
    \code{get\_yukawa\_eta()} & all $\{\eta_i^F\}$\\
    \code{get\_vevs()} & $\{v \cos\beta , v \sin\beta e^{i\xi}\}$\\
    \code{get\_higgs\_treeLvl\_masses()} &  $\{m_{h_i}, m_{H^\pm}\}$\\
    \code{get\_largest\_diagonal\_rF()} & max$(\rho_{ii}^F)$\\
    \code{get\_largest\_nonDiagonal\_rF()} & max$(\rho_{i\neq j}^F)$\\
    \code{get\_largest\_lambda()} & max$(\lambda_i)$\\
    \code{get\_largest\_nonDiagonal\_lamF()} & max$(\lambda_{i\neq j}^F)$\\
    \code{get\_lamF\_element(FermionSector,i,j)} & $\lambda_{ij}^F$\\ 
    \code{get\_lamF(FermionSector)} & $\lambda^F$ \\
    \code{get\_oblique()} & $\{S,T,U\}$
  \end{tabular}
  \end{center}
  Note that the aligned parameters $a^F$ are only meaningful when the Yukawa matrices are diagonal, which may change during RG evolution.

  The \code{get\_largest\_nonDiagonal\_lamF()} and \code{get\_lamF\_element(FermionSector,i,j)} functions returns the $\rho^F$ Yukawa matrices in terms of the Cheng-Sher ansatz \cite{ChengSher} defined by $\lambda_{ij}^F\equiv v \sqrt{\frac{\rho^F_{ij}}{2 m_i m_j}}$, where \code{FermionSector} is either \code{UP}, \code{DOWN} or \code{LEPTON}.

  The oblique parameters $S$, $T$ and $U$ are calculated with the formulas in \mycite{Haber:2010bw}.

There are also a number of functions that print information to the console:
\begin{itemize}
    \item \code{print\_higgs\_masses()}: Prints the tree-level Higgs boson masses.
    \item \code{print\_fermion\_masses()}: Prints the tree-level fermion masses.
    \item \code{print\_potential()}: Displays a table showing the scalar potential parameters. Both the generic and the Higgs basis are shown.
    \item \code{print\_yukawa()}: Prints all the $\eta^{F,0}$, $\kappa^F$ and $\rho^F$ Yukawa matrices.
    \item \code{print\_CKM()}: Prints the CKM matrix.
    \item \code{print\_param\_gen()}: Prints the scalar potential in the generic basis as well as the $\eta^{F,0}$ Yukawa matrices.
    \item \code{print\_param\_compact()}: Prints the scalar potential in the compact basis.
    \item \code{print\_oblique()}: Prints the oblique parameters $S$, $T$ and $U$.
    \item \code{print\_param\_higgs()}: Prints the scalar potential in the Higgs basis as well as the $\kappa^F$ and $\rho^F$ Yukawa matrices.
    \item \code{print\_features()}: Prints checks showing whether the model is \CP~conserving, perturbative, unitary and stable, where either is true or false.
    It also prints the results of any RGE running that might have been performed.
  \end{itemize}

If one wants to calculate the results of performing a RG evolution of the 2HDM without updating its parameters, one can use
\begin{center}
  \begin{tabular}{l|c|c}
    Function & Description & Returns\\ 
    \hline
    \code{calc\_rgeResults()} & Calculates RG evolution & \code{void}
\end{tabular}
\end{center}
The results can be printed to the console with the \code{RgeModel}'s function \code{print\_rgeResults}.
 
 One can create SLHA-like files and setting \THDM~objects with 
 \begin{center}
  \begin{tabular}{l|c|c}
    Function & Description & Returns\\   
    \hline
    \code{write\_slha\_file(string file)} & Creates SLHA file & \code{void}\\
    \code{set\_from\_slha\_file(string file)} & Sets \THDM~from SLHA file & \code{true/false}
  \end{tabular}
\end{center}
These files contain all the information of the \THDM.

%%%%%%%%%%%%%%%%%%%%%%%%%%%%%%%%%%%%%%%%%%%%%%%%%%%%%%%%%%%%%%%%%%%%%%%%%%%%%%%%
\section{RG evolution summary}\label{RGEalgorithm}
%%%%%%%%%%%%%%%%%%%%%%%%%%%%%%%%%%%%%%%%%%%%%%%%%%%%%%%%%%%%%%%%%%%%%%%%%%%%%%%%

Here, we briefly describe the procedure that is used when evolving a 2HDM.
As an example, we initialize the 2HDM at the top mass scale and evolve upwards in energy\footnote{Note that evolving a \THDM~downwards in energy is also possible. One must then, however, fix the high scale boundary condition first in some way.}.

First one must initialize the SM parameters, \textit{e.g.} by feeding \THDM~with a \SM.
Then one must set the scalar potential and Yukawa sector\footnote{Alternatively, one can set the \THDM~from a SLHA file instead of performing the first few steps.}.
The options for the RG evolution are specified by a \code{RgeConfig}, which is given to the \THDM~with \code{set\_rgeConfig}.
After that, the \THDM~can be evolved in energy with \code{evolve()}.
During the RG evolution, the following is happening:
\begin{itemize}
  \item At each intermediate step as specified in the \code{RgeConfig}, perturbativity, unitarity and stability are checked.
  \item The parameters are evolved in the generic basis, but the other bases are calculated with the $\mu$-dependent $\tan\beta$ and $\xi$ at each step.
  \item If \code{dataOutput=true}, the \THDM~stores the parameters as a function of $\mu$ in text files in \file{output/"evolutionName"}. If \code{GNUPLOT} is enabled, it also creates simple plots of the running parameters afterwards.
  \item When the RG evolution stops depends on the settings in \code{RgeConfig}. By default, it stops when perturbativity is violated; which with the default settings is very close to a Landau pole.
\end{itemize}

%%%%%%%%%%%%%%%%%%%%%%%%%%%%%%%%%%%%%%%%%%%%%%%%%%%%%%%%%%%%%%%%%%%%%%%%%%%%%%%
\section{Extending 2HDME}\label{extension}
%%%%%%%%%%%%%%%%%%%%%%%%%%%%%%%%%%%%%%%%%%%%%%%%%%%%%%%%%%%%%%%%%%%%%%%%%%%%%%%

To use \THDME~to evolve another QFT model or some 2HDM with additional degrees of freedom, one can either create a new model that inherits RGE functionality from \code{RgeModel} or extend the \THDM~class.

As a pedagogical example, there is a minimalistic class that simply describes the evolution of the gauge coupling in quantum electrodynamics called \code{NewModel}.
A demonstration of it is provided in \file{src/demos/DemoNewModel}.
To create one's own model, the procedure is the following:
\begin{itemize}
  \item Store all the models parameters as member variable of a class that inherits from \code{RgeModel}.
  All of the \code{virtual} functions must be overwritten in the derived class, see \code{NewModel} for an example.
  \item The member variables that should be evolved with RGEs should be transformed into an array with the \code{set\_y} function and the inverted transformation should be performed by \code{set\_model\_from\_y}.
  \item The variables in the array are evolved according to the RGEs that are contained in two functions: \code{rgeFuncNewModel\_1loop} and \code{rgeFuncNewModel\_2loop}.
   \item The function \code{rge\_update} should update the class's member variables and, as previously mentioned, the most minimalistic version of such a function is provided in \code{NewModel}.
\end{itemize}
The same procedure is applicable if one wants to add additional parameters to the \THDM~class.

Implementing higher order RGEs can also be done fairly easy.
The RGEs are divided into different functions in order to speed up evolutions of scenarios where one only needs the 1-loop order.
The 2-loop order functions in \file{RGE.h} do however compute all the loop contributions.
So if one wants to add additional loop orders, \textit{e.g.} the 3-loop contributions to the scalar potential in \mycite{Bednyakov:2018cmx}, simply modify \code{rgeFuncThdm\_2loop} to also compute these additional terms in the RGEs.

%%%%%%%%%%%%%%%%%%%%%%%%%%%%%%%%%%%%%%%%%%%%%%%%%%%%%%%%%%%%%%%%%%%%%%%%%%%%%%%%%
\section{Comparison to other software}\label{comparison}
%%%%%%%%%%%%%%%%%%%%%%%%%%%%%%%%%%%%%%%%%%%%%%%%%%%%%%%%%%%%%%%%%%%%%%%%%%%%%%%%%

Currently there is a number of different software programs that can be used to calculate RGEs of QFT models.
To generate RGEs up to 2-loop order, one can use \code{PyR@TE} \cite{Lyonnet:2013dna} and \code{SARAH} \cite{Staub:2013tta,Staub:2008uz}; they provide RGEs in \LaTeX or python/Mathematica code\footnote{\code{SARAH} also provides the option to export model files that can be used with \code{SPheno} \cite{Porod:2003um,Porod:2011nf} that can perform numerical evolution of the models.}.
The results of these programs do agree with the RGEs that we have derived from \mycite{Luo:2002ti} in the \Zsym symmetric 2HDM case.
However, we have not been able to generate consistent 2-loop RGEs in the general 2HDM with complex parameters with these programs; making a complete comparison difficult.

\THDME~serves a function as an API that provides fast numerical evolution of the general 2HDM in \code{C++}.
The user does not have to go through the trouble of creating a 2HDM model from scratch.
In addition, \THDME~is easily modified and self-contained\footnote{Except for linear algebra libraries such as \code{gsl} and \code{Eigen}.} and any numerical calculation should be completely transparent in an investigation of the source code.

%%%%%%%%%%%%%%%%%%%%%%%%%%%%%%%%%%%%%%%%%%%%%%%%%%%%%%%%%%%%%%%%%%%%%%%%%%%%%%%%%
\section{Conclusion}\label{conclusion}
%%%%%%%%%%%%%%%%%%%%%%%%%%%%%%%%%%%%%%%%%%%%%%%%%%%%%%%%%%%%%%%%%%%%%%%%%%%%%%%%%

We have described the \code{C++} program \THDME, which provides an API for evolving a general, potentially \CP-violating, 2HDM in renormalization scale by numerically solving its 2-loop RGEs.
Its main feature is the class \code{THDM} that represents a 2HDM object which is easily manipulated; with several parameterizations of the scalar potential available.
Furthermore, tree-level constraints of perturbativity, unitarity and stability are implemented; as well as calculations of the oblique parameters $S$, $T$ and $U$.

%%%%%%%%%%%%%%%%%%%%%%%%%%%%%%%%%%%%%%%%%%%%%%%%%%%%%%%%%%%%%%%%%%%%%%%%%%%%%%%%
\section*{Acknowledgments}

\THDME~originated from an extension of \code{2HDMC} \cite{Eriksson:2009ws} in collaboration with Johan Rathsman, who also provided useful feedback on this manuscript.

This work is supported in part by the Swedish Research Council grants contract numbers 621-2013-4287 and 2016-05996 and by the European Research Council (ERC) under the European Union’s Horizon 2020 research and innovation programme (grant agreement No 668679).

%%%%%%%%%%%%%%%%%%%%%%%%%%%%%%%%%%%%%%%%%%%%%%%%%%%%%%%%%%%%%%%%%%%%%%%%%%%%%%%%

\newpage

%%%%%%%%%%%%%%%%%%%%%%%%%%%%%%%%%%%%%%%%%%%%%%%%%%%%%%%%%%%%%%%%%%%%%%%%%%%%%%%%
\appendix
%%%%%%%%%%%%%%%%%%%%%%%%%%%%%%%%%%%%%%%%%%%%%%%%%%%%%%%%%%%%%%%%%%%%%%%%%%%%%%%%

%%%%%%%%%%%%%%%%%%%%%%%%%%%%%%%%%%%%%%%%%%%%%%%%%%%%%%%%%%%%%%%%%%%%%%%%%%%%%%%%%
\section{Installation instructions}\label{installation}
%%%%%%%%%%%%%%%%%%%%%%%%%%%%%%%%%%%%%%%%%%%%%%%%%%%%%%%%%%%%%%%%%%%%%%%%%%%%%%%%%

\subsection*{Source code}

The source code is available at \url{https://github.com/jojelen/2HDME}.

\subsection*{Dependencies}

The \THDME~requires the following to be installed:
\begin{itemize}
   \item A \code{C++11} compiler such as \code{gcc}.
   \item 2HDME requires the library Eigen \cite{eigenweb} to perform linear algebra operations. 
   If you have installed it, but the compiler does not find it, you can add the path to its directory with CFLAGS+=-I/.../eigen3 in the Makefile.

    \item To solve the RGEs, 2HDME uses the GNU GSL library \cite{GSL}, which is usually included in GNU/Linux distributions. See https://www.gnu.org/software/gsl/ for details.

\end{itemize}
\subsection*{Additional dependencies}

These dependencies are optional and can be enabled/disabled by commenting the
relevant lines in the Makefile:
\begin{itemize}
 \item The \THDME~can automatically create simple plots of the RG running of the
parameters with the help of \code{GNUPLOT}, see \file{http://www.gnuplot.info/} for details.

\end{itemize}

\subsection*{Compilation}

First, make sure that all the requirements are properly installed.
One might need to configure \file{Makefile} to link all the libraries if they are not installed
in the usual location. 
After that, one can proceed to compile: In terminal, type
\begin{mdframed}[style=MyFrame,userdefinedwidth=10em]
\code{make}
\end{mdframed}
in the main directory that contains the \file{Makefile}. 
Please note that the RGEs in \file{RGE.cpp} are not written in an
optimal form. 
However, the compiler with optimization level -03 will take care
of this. 
This takes some time and compiling \file{RGE.cpp} takes roughly 10 min on a laptop.

\subsection*{Run Demo}

To see that everything works, the demo in \sec{demo} is included.
The demo evolves a \Zsym symmetric \CP~conserved 2HDM from the top mass scale to the Planck scale. The source code is located in \file{src/demos/DemoRGE.cpp}.
To run it, type
\begin{mdframed}[style=MyFrame,userdefinedwidth=10em]
\code{bin/DemoRGE}
\end{mdframed}
in the terminal.
If the \code{GNUPLOT} functionality hasn't been disabled (by commenting out lines in
the Makefile), plots of the parameters should have been created in
\file{output/DemoRGE/plots}.

%%%%%%%%%%%%%%%%%%%%%%%%%%%%%%%%%%%%%%%%%%%%%%%%%%%%%%%%%%%%%%%%%%%%%%%%%%%%%%%%
\section{Bases of 2HDM's potential}\label{bases}
%%%%%%%%%%%%%%%%%%%%%%%%%%%%%%%%%%%%%%%%%%%%%%%%%%%%%%%%%%%%%%%%%%%%%%%%%%%%%%%%

There are four bases for the most general scalar potential of the 2HDM implemented in \THDME~as well as one basis that describes the \CP~conserved 2HDM with a softly broken \Zsym symmetry.
The base \code{struct} for these bases is \code{ThdmBasis}, which has the member variables that defines the VEV in \eq{vev}, i.e.\ $\beta$ and $\xi$.

More details about these bases and their relations to each other can be found in \mycites{Davidson:2005cw,Haber:2006ue}. 
Note that \THDM~works in the generic basis internally.

\subsection*{\code{Base\_generic}}

The generic basis of 2HDM is described in \sec{genericBasis}.
It consists of the additional parameters:
\begin{center}
  \begin{tabular}{c|c}
    Parameter & \code{Base\_generic}\\
    \hline
    $m_{11}^2$ & \code{M112}\\
    $m_{22}^2$ & \code{M222}\\
    $m_{12}^2$ & \code{M12}\\
    $\lambda_1$ & \code{Lambda1}\\
    $\lambda_2$ & \code{Lambda2}\\
    $\lambda_3$ & \code{Lambda3}\\
    $\lambda_4$ & \code{Lambda4}\\
    $\lambda_5$ & \code{Lambda5}\\
    $\lambda_6$ & \code{Lambda6}\\
    $\lambda_7$ & \code{Lambda7}
  \end{tabular}
\end{center}
A generic basis can be converted to other bases with the functions:
\begin{center}
  \begin{tabular}{l|l}
     & Returns\\
    \hline
    \code{convert\_to\_compact()} & \code{Base\_compact}\\
    \code{convert\_to\_higgs()} & \code{Base\_higgs}\\
    \code{convert\_to\_invariant(double v2)} & \code{Base\_invariant}
  \end{tabular}
\end{center}

\subsection*{\code{Base\_compact}}

The compact basis is defined by
\begin{align}\label{eq:CompactPotential}
	-\mathcal{L}_V = Y_{a\bar{b}}\Phi_{\bar{a}}^\dagger \Phi_b + \frac{1}{2} Z_{a\bar{b}c\bar{d}}(\Phi_{\bar{a}}^\dagger\Phi_b)(\Phi_{\bar{c}}^\dagger \Phi_d).
\end{align}
All these parameters are stored as complex numbers in \code{Base\_compact}:
\begin{center}
  \begin{tabular}{c|c}
    Parameter & \code{Base\_compact}\\
    \hline
    $Y_{a\bar{b}}$ & \code{Y[2][2]}\\
    $Z_{a\bar{b}c\bar{d}}$ & \code{Z[2][2][2][2]}
  \end{tabular}
  \end{center}

\subsection*{\code{Base\_higgs}}

The Higgs basis is defined by the basis where only one scalar doublet acquires a VEV, 
\begin{align}
  \braket{H_1} =  \frac{v}{\sqrt{2}} \left( \begin{array}{c} 0 \\ 1 \end{array} \right), && \braket{H_2} = 0.
\end{align}
The Lagrangian takes the form
\begin{align}\label{eq:HiggsPotential}
	-\mathcal{L}_V =~& Y_1 H_1^\dagger H_1 + Y_2 H_2^\dagger H_2 + \left(Y_3H_1^\dagger H_2+h.c.\right) + \frac{1}{2}Z_1(H_1^\dagger H_1)^2 + \frac{1}{2}Z_2(H_2^\dagger H_2)^2\nn
	&+ \frac{1}{2}Z_3(H_1^\dagger H_1)(H_2^\dagger H_2)+ \frac{1}{2}Z_4(H_1^\dagger H_2)(H_2^\dagger H_1)\nn
	&+\left\{\frac{1}{2}Z_5(H_1^\dagger H_2)^2 + \left[Z_6(H_1^\dagger H_1) + Z_7(H_2^\dagger H_2)\right]H_1^\dagger H_2 + \text{h.c.}\right\}.
\end{align}
The $Y_1,~Y_2,~Z_{1,2,3,4}$ are invariants under a Higgs flavor U(2) transformation, while $Y_3,~Z_{5,6,7}$ are pseudoinvariants that transform as
\begin{align}
	\{Y_3,Z_6,Z_7\} \rightarrow & (\det U)^{-1}\{Y_3,Z_6,Z_7\},\\
	Z_5 \rightarrow & (\det U)^{-2}Z_5.
\end{align}
This effectively means that the Higgs basis is unique up to a rephasing of the $H_2$ field. 
The parameters of \code{Base\_higgs} are:
\begin{center}
  \begin{tabular}{c|c}
    Parameter & \code{Base\_higgs}\\
    \hline
    $Y_1$ & \code{Y1}\\
    $Y_2$ & \code{Y2}\\
    $Y_3$ & \code{Y3}\\
    $Z_1$ & \code{Z1}\\
    $Z_2$ & \code{Z2}\\
    $Z_3$ & \code{Z3}\\
    $Z_4$ & \code{Z4}\\
    $Z_5$ & \code{Z5}\\
    $Z_6$ & \code{Z6}\\
    $Z_7$ & \code{Z7}
  \end{tabular}
\end{center}
And it can be transformed to other bases with the functions:
\begin{center}
  \begin{tabular}{l|l}
     & Returns\\
    \hline
    \code{convert\_to\_generic()} & \code{Base\_generic}\\
    \code{convert\_to\_compact()} & \code{Base\_compact}\\
    \code{convert\_to\_invariant(double v2)} & \code{Base\_invariant}
  \end{tabular}
\end{center}

\subsection*{\code{Base\_invariant}}

The invariant basis describes the general 2HDM with only Higgs flavor U(2) invariant quantities.

Four invariant quantities are the tree-level masses of the Higgs bosons.
The charged Higgs boson mass is given by
\begin{align}
	m_{H^\pm}^2 = Y_2 + \frac{1}{2} Z_3 v^2
\end{align} 
and the three neutral Higgs bosons' masses are given by the mass matrix in the Higgs basis,
\begin{align}\small
	\mathcal{M} \equiv v^2 \left( \begin{array}{ccc} Z_1 & \real{Z_6} & -\imag{Z_6} \\
												\real{Z_6} & \frac{1}{2}\left[Z_3 + Z_4 + \real{Z_5}\right] + Y_2/v^2 & -\frac{1}{2}\imag{Z_5} \\
												-\imag{Z_6}	&	-\frac{1}{2}\imag{Z_5} & \frac{1}{2}\left[Z_3 + Z_4 -\real{Z_5}\right] + Y_2/v^2 \end{array} \right).
\end{align}
This mass matrix can be diagonalized with the rotation matrix
\begin{align}\small
	R \equiv \left( \begin{array}{ccc} c_{13}c_{12} & -c_{23}s_{12}-c_{12}s_{13}s_{23} & -c_{12}c_{23}s_{13} + s_{12}s_{23} \\
												c_{13}s_{12} & c_{12}c_{23} - s_{12}s_{13}s_{23} & -c_{23}s_{12}s_{13}-c_{12}s_{23} \\
												s_{13}	&	c_{13}s_{23} & c_{13}c_{23} \end{array} \right),
\end{align}
to produce $\mathcal{M}_D = R \mathcal{M} R^T = \text{diag}( m_{h_1}^2, m_{h_2}^2, m_{h_3}^2)$.
We will, without loss of generality, assume ordered masses, $m_{h_1} < m_{h_2} < m_{h_3}$.
The eigenvalues of the mass matrix are invariant under Higgs flavor transformations, even though the matrix elements are not.
Consequently, the rotation matrix is not invariant either.
While the angles $\theta_{12}$ and $\theta_{13}$ are invariant, $\theta_{23}$ changes so that $e^{i\theta_{23}} \rightarrow (\det U)e^{i\theta_{23}}$ is a pseudo-invariant quantity \cite{Haber:2006ue}.
In \mycite{Haber:2006ue}, they define a U(2) invariant mass matrix,
\begin{align}\small
	\tilde{\mathcal{M}} \equiv v^2 \left( \begin{array}{ccc} Z_1 & \real{Z_6e^{-i\theta_{23}}} & -\imag{Z_6e^{-i\theta_{23}}} \\
												\real{Z_6 e^{-i\theta_{23}}} & \real{Z_5e^{-2i\theta_{23}}} + A^2/v^2 & -\frac{1}{2}\imag{Z_5e^{-2i\theta_{23}}} \\
												-\imag{Z_6e^{-i\theta_{23}}}	&	-\frac{1}{2}\imag{Z_5e^{-2i\theta_{23}}} &  A^2/v^2 \end{array} \right),
\end{align}
where $A^2 \equiv Y_2 + \frac{1}{2}\left[Z_3+Z_4-\real{Z_5e^{-2i\theta_{23}}}\right]v^2$.
This matrix is  diagonalized with the rotation matrix
\begin{align}\small
	\tilde{R} \equiv \left( \begin{array}{ccc} c_{12}c_{13} & -s_{12} & -c_{12}s_{13}  \\
												c_{13}s_{12} & c_{12}  & -s_{12}s_{13} \\
												s_{13}	&	0 & c_{13} \end{array} \right),
\end{align}
such that $\mathcal{M}_D =\tilde{R} \tilde{\mathcal{M}} \tilde{R}^T$.
The angles lie in the range $-\pi/2 \leq \theta_{12}, \theta_{13} < \pi/2$ in general.

Now, we can construct a U(2) invariant basis with the 8 real parameters $\{m_{h_i}\}$, $m_{H^\pm}$, $\theta_{12}$, $\theta_{13}$, $Z_2$, $Z_3$ and the complex parameter $Z_7e^{-i\theta_{23}}$:
\begin{center}
  \begin{tabular}{c|c|c}
    Parameter & Description & \code{Base\_invariant}\\
    \hline
			$\{m_{h_1}, m_{h_2}, m_{h_3} \}$ 	& Ordered neutral Higgs boson masses & \code{mh[3]}\\
			$m_{H^\pm}$ & Charged Higgs boson mass & \code{mHc}\\
			$s_{12}\in [-1,1]$ & Mixing angle of neutral Higgs mass matrix & \code{s12}\\
			$c_{13}\in [0,1]$ & Mixing angle of neutral Higgs mass matrix & \code{c13}\\
			$\{Z_2, Z_3\}$ & Real quartic couplings & \code{Z2,Z3}\\
			$Z_{7}e^{-i\theta_{23}}$ & Complex quartic coupling & \code{Z7inv}\\
    \end{tabular}
  \end{center}

The invariant basis can be converted to other bases with
\begin{center}
  \begin{tabular}{l|l}
     & Returns\\
    \hline
    \code{convert\_to\_generic(double v2)} & \code{Base\_generic}\\
    \code{convert\_to\_compact(double v2)} & \code{Base\_compact}\\
    \code{convert\_to\_higgs(double v2)} & \code{Base\_higgs}
  \end{tabular}
\end{center}

\subsection*{\code{Base\_hybrid}}

The hybrid basis presented in \mycite{Haber:2015pua} is describing a \CP~conserved 2HDM with a softly broken \Zsym symmetry.
It consists of a combination of tree-level masses and quartic couplings:
\begin{center}
  \begin{tabular}{c|c|c}
    Parameter & Description & \code{Base\_invariant}\\
    \hline
			$m_{h}$ & Lightest \CP~even Higgs boson & \code{mh}\\
			$m_{H}$ & Heaviest \CP~even Higgs boson & \code{mH}\\
			$\cos(\beta-\alpha)$ & Mixing angle of \CP~even Higgs mass matrix & \code{cba}\\
			$\tan\beta$ & Ratio of Higgs VEVs in generic basis & \code{tanb}\\
			$\{Z_4, Z_5, Z_7\}$ & Real quartic couplings & \code{Z4,Z5,Z7}
    \end{tabular}
  \end{center}
and can be converted to the general bases with
\begin{center}
  \begin{tabular}{l|l}
     & Returns\\
    \hline
    \code{convert\_to\_generic(double v2)} & \code{Base\_generic}\\
    \code{convert\_to\_higgs(double v2)} & \code{Base\_higgs}\\
    \code{convert\_to\_invariant(double v2)} & \code{Base\_invariant}
  \end{tabular}
\end{center}

%%%%%%%%%%%%%%%%%%%%%%%%%%%%%%%%%%%%%%%%%%%%%%%%%%%%%%%%%%%%%%%%%%%%%%%%%%%%%%%%%
\section{SM input}\label{SMinput}
%%%%%%%%%%%%%%%%%%%%%%%%%%%%%%%%%%%%%%%%%%%%%%%%%%%%%%%%%%%%%%%%%%%%%%%%%%%%%%%%%

The \SM~is defined at the top quark mass scale, $M_t=173.34$ GeV \cite{ATLAS:2014wva}.
See \sec{SM} for instructions of how to create a \SM~object at another renormalization scale. 
At construction, we use the following input to fix its parameters:
\begin{itemize}
  \item The SM Higgs VEV is taken to be $v= (\sqrt{2} G_F)^{-1/2} = 246.21971$ GeV \cite{Buttazzo:2013uya}.
  
	\item	The fermion masses are used to fix the Yukawa matrix elements in the fermion mass eigenbasis. We use the ones from \mycite{Xing:2007fb}:
	 \begin{align}
	 m_u =~& 1.22\text{ MeV}, &  m_c =~& 0.590\text{ GeV}, &  m_t =~& 162.2 \text{ GeV},\nn
     m_d =~& 2.76\text{ MeV}, &  m_s =~& 52\text{ MeV}, & m_b =~&  2.75\text{ GeV},\nn
     m_e =~& 0.485289396\text{ MeV}, & m_\mu =~& 0.1024673155\text{ GeV}, & m_\tau =~&  1.74215 \text{ GeV}.
     \end{align}
	
	\item Gauge couplings from \mycite{Buttazzo:2013uya}:
	\begin{align}
		g_1 =~& 0.3583\nn
		g_2 =~& 0.64779\nn
		g_3 =~& 1.1666
	\end{align}
	for U(1)$_Y$, SU(2)$_W$ and SU(3)$_c$ respectively.

	\item For the CKM matrix, we use the standard parametrization
	\begin{align}\label{eq:CKM}
		V_{CKM} = \left( \begin{array}{ccc} c_{12}c_{13} & s_{12}c_{13} & s_{13}e^{-i\delta} \\ 
		-s_{12}c_{23} - c_{12}s_{23}s_{13}e^{i\delta} & c_{12}c_{23}-s_{12}s_{23}s_{13}e^{i\delta} & s_{23}c_{13} \\ 
		s_{12}s_{23} - c_{12}c_{23}s_{13}e^{i\delta} & -c_{12}s_{23}-s_{12}c_{23}s_{13}e^{i\delta}& c_{23}c_{13}\end{array}\right),
	\end{align}
	where the angles in terms of the Wolfenstein parameters are
	\begin{align}
		s_{12} =~& \lambda,\nn
		s_{23} =~& A\lambda^2,\nn
		s_{13}e^{i\delta} =~& \frac{A\lambda^3(\bar{\rho} + i\bar{\eta})\sqrt{1-A^2\lambda^4}}{\sqrt{1-\lambda^2}\left[1-A^2\lambda^4(\bar{\rho}+i\bar{\eta})\right]}.
	\end{align}
	The numerical values 
	\begin{align}
		\lambda =~& 0.22453,\nn
        A =~& 0.836,\nn
        \bar{\rho} =~& 0.122,\nn
        \bar{\eta} =~& 0.355,
	\end{align}
	are extracted from the PDG \cite{PDG}. 
\end{itemize}

%%%%%%%%%%%%%%%%%%%%%%%%%%%%%%%%%%%%%%%%%%%%%%%%%%%%%%%%%%%%%%%%%%%%%%%%%%%%%%%%
\section{Tree-level unitarity conditions} \label{unitarity}
%%%%%%%%%%%%%%%%%%%%%%%%%%%%%%%%%%%%%%%%%%%%%%%%%%%%%%%%%%%%%%%%%%%%%%%%%%%%%%%%

The tree-level unitarity conditions for a general 2HDM have been worked out in \mycite{Ginzburg:2005dt}.
There, they work out the following scattering matrices:
\begin{align}
	\Lambda_{21} \equiv~ & \left( \begin{array}{ccc} \lambda_1 & \lambda_5 & \sqrt{2}\lambda_6 \\
										\lambda_5^* & \lambda_2 & \sqrt{2}\lambda_7^* \\
										\sqrt{2}\lambda_6^* & \sqrt{2}\lambda_7 & \lambda_3 + \lambda_4 \end{array}\right),\\
	\Lambda_{20} \equiv~ & \lambda_3 - \lambda_4,\\
	\Lambda_{01} \equiv~ & \left( \begin{array}{cccc} \lambda_1 & \lambda_4 & \lambda_6 & \lambda_6^* \\
										\lambda_4 & \lambda_2 & \lambda_7 & \lambda_7^* \\
										\lambda_6^* & \lambda_7^* & \lambda_3 & \lambda_5^* \\
										\lambda_6 & \lambda_7 & \lambda_5 & \lambda_3 \end{array}\right),\\	
	\Lambda_{00} \equiv~ & \left( \begin{array}{cccc} 3\lambda_1 & 2\lambda_3 + \lambda_4 & 3\lambda_6 & 3\lambda_6^* \\
										2\lambda_3 + \lambda_4 & 3\lambda_2 & 3\lambda_7 & 3\lambda_7^* \\
										3\lambda_6^* & 3\lambda_7^* & \lambda_3 + 2\lambda_4 & 3\lambda_5^* \\
										3\lambda_6 & 3\lambda_7 & 3\lambda_5 & \lambda_3 + 2\lambda_4 \end{array}\right).	
\end{align}
In the end, the unitarity constraint put upper limits on the absolute value of the eigenvalues, $\Lambda_i$, of these matrices,
\begin{align}
	\abs{\Lambda_i} < 8\pi.
\end{align}

%%%%%%%%%%%%%%%%%%%%%%%%%%%%%%%%%%%%%%%%%%%%%%%%%%%%%%%%%%%%%%%%%%%%%%%%%%%%%%%%
\section{Tree-level stability of the scalar potential} \label{stability}
%%%%%%%%%%%%%%%%%%%%%%%%%%%%%%%%%%%%%%%%%%%%%%%%%%%%%%%%%%%%%%%%%%%%%%%%%%%%%%%%

Here, we give the conditions for the scalar potential to be bounded from below, as worked out in \mycite{Ivanov:2006yq, Ivanov:2007de}.

When working out these conditions, \mycite{Ivanov:2006yq, Ivanov:2007de} constructed a Minkowskian formalism of the 2HDM that uses gauge-invariant field bilinears,
\begin{align}
	r^0 \equiv~& \Phi_1^\dagger \Phi_1 + \Phi_2^\dagger \Phi_2,\\
	r^1 \equiv~& 2\real{\Phi_1^\dagger \Phi_2},\\
	r^2 \equiv~& 2\imag{\Phi_1^\dagger \Phi_2},\\
	r^3 \equiv~& \Phi_1^\dagger \Phi_1 - \Phi_2^\dagger \Phi_2.
\end{align}
These can be used to create a four-vector $r^\mu = (r^0, \vec{r})$; where one can raise and lower the indices as usual with the flat Minkowski metric $\eta^{\mu\nu} = \text{diag}(1,-1,-1,-1)$.
In this formalism, the scalar potential is conveniently written as
\begin{align}
	V = -M_\mu r^\mu + \frac{1}{2} r^\mu \Lambda_\mu^\nu r_\nu,
\end{align}
where 
\begin{align}
	M_\mu = \left( -\frac{1}{2}(Y_1+Y_2), \real{Y_3}, -\imag{Y_3}, -\frac{1}{2}(Y_1-Y_2)\right)
\end{align}
and
\begin{align}\small
	\Lambda_\mu^\nu = \frac{1}{2} \left( \begin{array}{cccc} \frac{1}{2}(Z_1 + Z_2) + Z_3 & -\real{Z_6 + Z_7} & \imag{Z_6 + Z_7} & -\frac{1}{2}(Z_1-Z_2) \\
	\real{Z_6 + Z_7} & -Z_4-\real{Z_5} & \imag{Z_5} & -\real{Z_6-Z_7}\\
	-\imag{Z_6+Z_7} & \imag{Z_5} & -Z_4 +\real{Z_5} & \imag{Z_6-Z_7}\\
	\frac{1}{2}(Z_1-Z_2) & -\real{Z_6-Z_7} & \imag{Z_6 - Z_7} & -\frac{1}{2}(Z_1+Z_2)+Z_3\end{array}\right).
\end{align}
The scalar potential is bounded from below \textit{if and only if} all of the below requirements are fulfilled:
\begin{itemize}
	\item All the eigenvalues of $\Lambda_\mu^\nu$ are real.
	\item There exists a largest eigenvalue that is positive, $\Lambda_0>\{\Lambda_1, \Lambda_2, \Lambda_3\}$ and $\Lambda_0>0$.
	\item There exist four linearly independent eigenvectors; one $V^{(a)}$ for each eigenvalue $\Lambda_a$.
	\item The eigenvector to the largest eigenvalue is timelike, while the others are spacelike,
	\begin{align}
		V^{(0)}\cdot V^{(0)} =~& \left(V^{(0)}_0\right)^2 - \sum_{i=1}^3\left(\vec{V}^{(0)}_i\right)^2>0,\\
		V^{(i)}\cdot V^{(i)} =~& \left(V^{(i)}_0\right)^2 - \sum_{j=1}^3\left(\vec{V}^{(i)}_j\right)^2<0.
	\end{align}
\end{itemize} 

%%%%%%%%%%%%%%%%%%%%%%%%%%%%%%%%%%%%%%%%%%%%%%%%%%%%%%%%%%%%%%%%%%%%%%%%%%%%%%%%
\section{Example of output} \label{DemoOutput}
%%%%%%%%%%%%%%%%%%%%%%%%%%%%%%%%%%%%%%%%%%%%%%%%%%%%%%%%%%%%%%%%%%%%%%%%%%%%%%%%

The \file{DemoRGE} program evolves a 2HDM from the top mass scale to the Planck scale, as explained in \sec{demo}.
It produces a SLHA text file that contains different blocks with the numerical values of parameters and test results.
%  and configurations for \code{SPheno}.
Here, we list some of the blocks for the 2HDM after RG evolution.

The first two blocks contain the renormalization scale where all parameters are defined, as well as potential violations of tree-level tests and the results from any RG evolution that may have been performed of the model:
\begin{verbatim}
Block THDME   # Features at current scale
 0     1.00000000e+18   # Renormalization scale
 1   1   # Yukawa symmetry (0=none or Type 1-4)
 2   1   # CP conserving (0=false, 1=true)
 3   1   # Perturbative (0=false, 1=true)
 4   1   # Unitary, tree-lvl (0=false, 1=true)
 5   1   # Stable, tree-lvl  (0=false, 1=true)
 6   1   # Stable, tree-lvl and z2sym  (0=false, 1=true)
Block RGE   # Results of RGE evolution
 0     1.73340000e+02   # Start scale
 1     1.00000000e+18   # Finish scale (same as current renorm scale)
 2    -1.00000000e+00   # Perturbativity breakdown scale (-1 = no violation)
 3    -1.00000000e+00   # Unitarity breakdown scale (-1 = no violation)
 4    -1.00000000e+00   # Stability breakdown scale (-1 = no violation)
\end{verbatim}

The scalar potential parameters are stored in the following blocks:
\begin{verbatim}
Block MIJ2      # Mij^2
 1     5.36216563e+05   # M112
 2     4.48291729e+04   # M222
Block MINPAR      # Input parameters (real part)
 1     4.11882166e+00   # Lambda1Input
 2     1.74519531e-01   # Lambda2Input
 3     6.65974644e-02   # Lambda3Input
 4     6.03919222e-01   # Lambda4Input
 5    -6.36125269e-01   # re(Lambda5Input)
 6     0.00000000e+00   # re(Lambda6Input)
 7     0.00000000e+00   # re(Lambda7Input)
 9     4.92420443e+04   # re(M12input)
 10    1.85559414e+00   # TanBeta
Block IMMINPAR      # Input parameters (imaginary part)
 5     1.89412625e-19   # im(Lambda5Input)
 6     0.00000000e+00   # im(Lambda6Input)
 7     0.00000000e+00   # im(Lambda7Input)
 9     4.36640351e-15   # im(M12input)
\end{verbatim}

Also the VEVs and gauge couplings have blocks:
\begin{verbatim}
Block HMIXIN       #  
 102     1.11500464e+02   # re(v1)
 103     2.06899607e+02   # re(v2)
Block IMHMIXIN       #  
 102     0.00000000e+00   # im(v1)
 103     2.16540428e-27   # im(v2)
Block GAUGEIN       #  
 1     4.69220242e-01   # g1
 2     5.17295774e-01   # g2
 3     5.00709726e-01   # g3
\end{verbatim}

All six $\eta^{F,0}_a$ Yukawa matrices are stored in two blocks each, one for the real and one for the imaginary part.
For example, one of these look like:
\begin{verbatim}
Block ETA2UIN    # eta2U Yukawa matrix 
 1 1     2.98177234e-06   # eta2U(1,1)
 1 2    -7.16532643e-12   # eta2U(1,2)
 1 3    -2.00521524e-08   # eta2U(1,3)
 2 1    -1.48164372e-14   # eta2U(2,1)
 2 2     1.44200704e-03   # eta2U(2,2)
 2 3    -6.98535296e-07   # eta2U(2,3)
 3 1    -1.51464466e-13   # eta2U(3,1)
 3 2    -2.55170410e-09   # eta2U(3,2)
 3 3     4.61876270e-01   # eta2U(3,3)
\end{verbatim}

% In addition to these, if \code{SPheno} is enabled, there are blocks called \code{SPhenoInput} and \code{DECAYOPTIONS} that are used as input to \code{SPheno} and we thus refer to their documentation for a description of each parameter.
\bibliography{2HDMbib}

\end{document}